\documentclass[twocolumn,showpacs,amsmath,amssymb]{revtex4}
\newcommand{\beq}{\begin{eqnarray}}
\newcommand{\eeq}{\end{eqnarray}}
\newcommand{\beqq}{\begin{eqnarray*}}
\newcommand{\eeqq}{\end{eqnarray*}}
\def\be{\begin{equation}}
\def\ee{\end{equation}}
\def\bee{\begin{equation*}}
\def\eee{\end{equation*}}
\def\been{\begin{enumerate}}
\def\enen{\end{enumerate}}
\def\beit{\begin{itemize}}
\def\enit{\end{itemize}}

\usepackage{graphicx}% Include figure files
\usepackage{dcolumn}% Align table columns on decimal point
\usepackage{bm}% bold math
\begin{document}

\title{Broadband chaos generated by an opto-electronic oscillator}

\author{Kristine E. Callan}
\author{Lucas Illing}
\altaffiliation[Current address: ]{Reed College, Dept. of Physics, Portland, OR 97202 USA}
\author{Zheng Gao}
\author{Daniel J. Gauthier}
\author{Eckehard Sch\"{o}ll}
\altaffiliation[Permanent address: ]{Technische Universit\"{a}t Berlin, Institut f\"{u}r Theoret. Physik, Hardenbergst. 36, 10623 Berlin, Germany}
\affiliation{Duke University, Department of Physics, Durham, North Carolina 27708 USA}

\date{\today}% It is always \today, today,

\begin{abstract}
We study an opto-electronic time-delay oscillator that displays high-speed chaotic behavior with a flat, broad power spectrum. The chaotic state coexists with a linearly-stable fixed point, which, when subjected to a finite-amplitude perturbation, loses stability initially via a periodic train of ultrafast pulses.  We derive approximate mappings that do an excellent job of capturing the observed instability.  The oscillator provides a simple device for fundamental studies of time-delay dynamical systems and can be used as a building block for ultra-wide-band sensor networks. 
\end{abstract}


\maketitle

%{\bf Intro}
%Big picture paragraph (why are people interested in time-delay systems?)
A deterministically chaotic system displays extreme sensitivity to initial conditions and the spectra of the fluctuating system variables are broadband. Yet, for typical chaotic devices, the power spectra often contain several sharp features that stand out above a broad background, which are often associated with weakly unstable periodic orbits that are part of the backbone of the strange attractor. The fact that the power spectra for typical chaotic devices are not featureless limits their application in ultra-wide-band (UWB) sensor networks \cite{UWB2} and in chaos-based ranging devices \cite{Lin2004}, for example.  

In this Letter, we describe an opto-electronic time-delay oscillator that displays high-speed chaos with an essentially featureless power spectrum.  The chaotic behavior coexists with a linearly stable quiescent state.  If the system starts in this state, a finite-size perturbation is needed to force the system to the chaotic state.  We show that a sufficiently large perturbation causes the system to produce an initially periodic train of ultrafast pulses whose spacing and amplitude becomes irregular for longer times.  Our observations are in good agreement with the predictions of a nonlinear stability analysis of the fixed point.

Our work has important implications for understanding the stability of general time-delay systems, for which coexisting states are common.  For example, the stability and noise sensitivity of opto-electronic microwave oscillators \cite{Chembo2008}, synchronized neuronal networks \cite{Vicente2008}, synthetic gene networks \cite{Weber2007}, and controlled chaotic systems \cite{Handbook,Dahms2008} may be adversely affected by the presence of a coexisting chaotic state. Our analysis predicts the amplitude of noise or externally applied perturbations that allow such systems to `sense' the coexisting strange attractor.

\begin{figure}[h,t,b]
\includegraphics[scale=1]{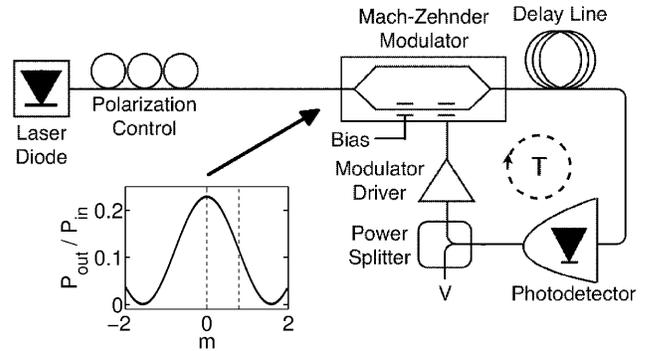}
\caption{\label{fig:setup} Experimental setup. Inset: Nonlinear transmission of the MZM (ratio of the output to input powers of the device) as a function of the dimensionless bias voltage $m$.}
\end{figure}

Our opto-electronic oscillator consists of a nonlinear element placed in a time-delay feedback loop and displays a variety of dynamical behaviors that depend on system parameters.  As shown in Fig.\ \ref{fig:setup}, the beam generated by a semiconductor laser (wavelength 1.55 $\mu$m) is injected into a single-mode optical fiber, passes through a polarization controller, and is injected into a Mach-Zehnder modulator (MZM).  The transmission of the MZM is a nonlinear function of the applied voltage, where we independently apply a time-dependent voltage to the radio-frequency (RF) port of the device (half-wave voltage $V_{\pi,RF}$=7.4 V) and a dc-voltage $V_{B}$ to bias it at any point on the transmission curve (half-wave voltage $V_{\pi,dc}$=7.7 V). Light exiting the modulator passes through an additional piece of single-mode fiber (length $\sim$5 m) serving as a delay line and is incident on a photodetector. Half of the resulting signal, denoted by $V$, is amplified by an inverting modulator driver (gain $g_{MD}=-22.6$) and fed back to the MZM via the ac-coupled input port. The modulator driver saturates at high voltage with saturation voltage $V_{sat}=9.7$ V. The other half of the signal is directed to a high-speed oscilloscope (8 GHz analog bandwidth, 40 GS/s sampling rate). The total delay of the feedback loop $T$=24.1 ns.

Similar opto-electronic oscillators have been studied previously, dating back to the seminal work of Ikeda \cite{Ikeda1982}. One distinguishing feature of our device is that the amplifier is ac-coupled so that feedback of low frequencies is suppressed.  Also, high frequencies are suppressed due to the finite response time of the photodetector and amplifier.  We find that the linear frequency response of the various components of the system is well described by a two-pole bandpass filter with a low- (high-) frequency cut-off $\omega_-=1.5\times10^5$ s$^{-1}$ ($\omega_+=7.5\times10^{10}$ s$^{-1}$), center frequency $\omega_0=\sqrt{\omega_-\omega_+}=1.1\times10^8$ s$^{-1}$, and bandwidth $\Delta=\omega_+-\omega_-=7.5\times10^{10}$ s$^{-1}$. The system is described in terms of a single time-delay integro-differential equation (or, equivalently, two coupled time-delay differential equations) as opposed to a single time-delay differential equation used by Ikeda. Such time-delay integro-differential equations display unique bifurcation properties \cite{LucasPD2005,Lucas2006} and new behaviors, such as chaotic breathers \cite{LargerPRL2005_2}.

Another important distinction of our work is that we bias the MZM at the maximum of the transmission curve ($m=\pi V_B/2V_{\pi,dc}=0$) shown in the inset of Fig.\ \ref{fig:setup}. As shown below, such a bias renders the quiescent state of the system linearly stable. Essentially all other research has focused on the case where the bias is set to the half-transmission point of the transmission curve ($m=\pi/4$), where the quiescent state is most linearly unstable.  Counter examples exist, such as the work of Meucci \textit{et al.} \cite{Meucci2002}, although they did not focus on the behavior discussed here.

The dynamics of our opto-electronic oscillator is described by the dimensionless coupled time-delay differential equations (DDEs) \cite{Lucas2006,LargerPRL2005_2}
\begin{align}
\dot x(s) & = -  x(s) - y(s) + c[x(s-\tau)] \label{eqn:model} \\ 
\dot y(s) & = \epsilon x(s). \label{eqn:dimensionless2}
\end{align}
Here, $c[x]=\gamma \cos^2\left(m + d\tanh x \right) - \gamma \cos^2m$ is the nonlinear delayed-feedback term, $x=g_{MD}V/V_{sat}$, the overdot denotes the derivative with respect to the dimensionless time $s=t\Delta$, $\gamma$ is the overall gain of the feedback loop and is proportional to the laser power, $\tau=T\Delta$, and $\epsilon=\omega_0^2/\Delta^2$ characterizes the bandpass filter.  Differing from Ref.\ \cite{LargerPRL2005_2}, we incude a hyperbolic tangent function in $c[x]$ to account for amplifier saturation, characterized by the parameter $d=\pi V_{sat}/2V_{\pi,RF}$.  In our experiments, three parameters are held fixed ($d=2.1$, $\tau=1820$, and $\epsilon=2.0 \times 10^{-6}$), while $\gamma$ can range from $0-5$ by adjusting the laser power with an attenuator and $m$ ranges from $-\pi/2$ to $\pi/2$.  For future reference, note that $x$ and $V$ have opposite signs because $g_{MD}<0$.

We first investigate the linear stability of one of the fixed points of Eqs.\ (\ref{eqn:model}) and (\ref{eqn:dimensionless2}) located at $(x^*,y^*)=(0,0)$, which is the quiescent state of the oscillator.  Linear stability analysis predicts that the fixed point is stable for small $\gamma$ and undergoes a Hopf bifurcation (a transition to an oscillatory behavior) at
\begin{align}
 \gamma_H=-\frac{b_{\pm}}{d \sin (2m)},\label{eqn:hopf}
\end{align}
where $b_{\pm}$ is a constant that depends on $\tau$ and $\epsilon$ and is approximately equal to $\pm1$ for our conditions. Clearly, the fixed point is linearly-stable for all $\gamma$ for $m=0$ where $\gamma_H$ diverges.

Experimentally, we find that the stability properties of the fixed point are much more complex than predicted by the linear analysis presented above.  In particular, near $m=0$, we find that finite-size perturbations destabilize the fixed-point for $\gamma < \gamma_H$, which can only be understood from a global (nonlinear) stability analysis of the model.

\begin{figure}[h,t,b]
\includegraphics[scale=1]{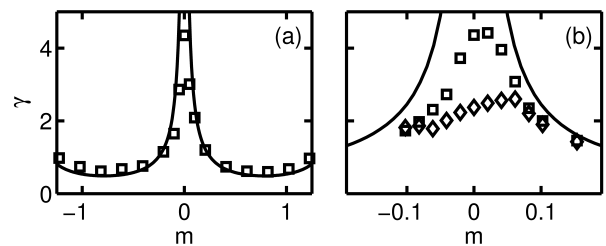}
\caption{\label{fig:gamma_th} Observed values of $\gamma$ for which the system transitions from steady-state to oscillatory or pulsing behavior as a function of $m$, with $\gamma_H$ superimposed (solid line). The squares in (a) and (b) indicate low experimental noise, while the diamonds in (b) indicate a higher level of noise.}
\end{figure}

In the experiments, we slowly increase $\gamma$ from zero until the fixed point $(x^*,y^*)=(0,0)$ is destabilized.  It is seen in Fig. \ref{fig:gamma_th}a that there is very good agreement between the predictions of the linear theory and experiment around $m=\pm \pi/4$ (the standard bias used in most previous experiments), but there is substantial disagreement in the vicinity of $m=0$ (see Fig.\ \ref{fig:gamma_th}b). At $m=0$, the system loses stability by transitioning directly to a broadband chaotic state at $\gamma=4.36$, undergoing transient pulsations en route to chaos.  The finite-size perturbations needed to destabilize the fixed point originate from noise in our system (\textit{e.g.}, laser relaxation oscillations, and detector dark and shot noise).

We add more noise to the system to measure its affect on the instability threshold. Insterting into the loop an erbium-doped fiber amplifier (EDFA) followed by an attenuator so that the total optical power injected into the oscillator is the same, we observe that the fluctuating part of the voltage $V$ increases by a factor of 2.3 (root-mean-square noise over a bandwidth from dc to 8 GHz). The open diamonds in Fig.\ \ref{fig:gamma_th} show that the instability threshold decreases for $|m|<0.1$ due to the increased noise.  There is also a pronounced asymmetry in the instability threshold about $m=0$.

\begin{figure}[h,t,b]
\includegraphics[scale=1]{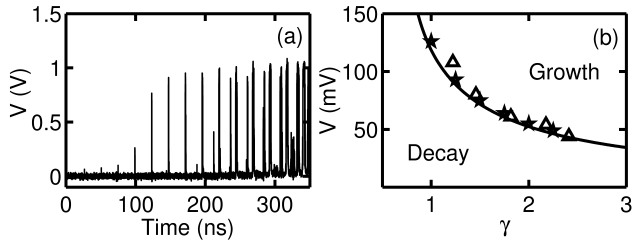}
\caption{\label{fig:comparison} (a) Experimentally observed transient behavior that results for $m=0$ and $\gamma=4.36$ when the system leaves the steady-state. The pulses have a FWHM $\sim 0.2$ ns and are separated by the time-delay $T$. (b) The critical pulse amplitude as a function of $\gamma$ in the experiment (triangles) and simulation (stars) with the unstable fixed point of the map superimposed.}
\end{figure}

In the vicinity of $m=0$, we observe that the system loses stability by generating a sequence of ultrashort pulses spaced initially by $T$ with a pulse duration (full width at half maximum) of $\sim 200$ ps. Typical transient behavior is shown in Fig.\ \ref{fig:comparison}a, where we have removed the EDFA from the setup and adjust $\gamma$ just above the instability threshold.  To more carefully study this transient behavior, we add an additional 3-dB power splitter to the feedback loop, lower $\gamma$ so that the system is in the quiescent state and inject 200-ps-long electrical pulses into the loop. For small pulse amplitude, the perturbation decays.  For sufficiently large pulse amplitude, we observe that the perturbation grows rapidly initially, levels off, and the waveform becomes more complex, similar to that shown in Fig.\ \ref{fig:comparison}a.  The open triangles shown in Fig.\ \ref{fig:comparison}b give the critical value of the pulse amplitude needed to destabilize the fixed point as a function of the feedback loop gain.

The instability boundaries shown in Fig.\ \ref{fig:gamma_th} and the transient pulsing behavior shown in Fig.\ \ref{fig:comparison} can be understood by asymptotic analysis of Eqs.\ (\ref{eqn:model}) and (\ref{eqn:dimensionless2}). Consider a finite-size perturbation to $x$ in the form of a short pulse of amplitude $-x_0$ centered at time $s=0$. Due to the smallness of $\epsilon \sim 10^{-6}$, $y$ responds extremely slowly to this pertubation, and hence we set $y=0$.  The perturbation propagates through the feedback loop and, near time $\tau$, the feedback term begins to grow from zero, corresponding to the presence of the pulse a time $\tau$ earlier. Approximately, this produces another pulse $c[x_0]$ in $x$. Thus, this second pulse at time $\tau$ will generate a third pulse at time $2\tau$, and so on, consistent with the behavior shown in Fig.\ \ref{fig:comparison}a.

From the asymptotic analysis, we derive a one-dimensional discrete map of the form 
\begin{equation}
x_{n+1}=c[x_n],\label{eqn:map}
\end{equation}
where $x_n$ can be thought of as the amplitude of a pulse at time $n\tau$. Further analysis reveals that there exists a continuous mapping of a time interal of length $\tau$ into the next future interval of length $\tau$.  It is given by
\begin{equation}
x_{s,s+\tau}=c[x_{s-\tau,s}],\label{eqn:cmap}
\end{equation}
where $x_{a,b}$ denotes the temporal evolution of $x$ over the interval $(a,b]$.  One should keep in mind, however, that these mappings only approximately predict the dynamics of the physical system, as reducing the coupled DDEs to a mapping erases all of the effects of the bandpass filter.

The discrete map (\ref{eqn:map}) has multiple fixed points, depending on the values of $\gamma$, $m$, and $d$.  Here, we focus on the case of $m=0$, where there are one or three fixed points for $d=2.1$.  We find that the fixed point $x_{s1}^*=0$ is always stable. It corresponds to the steady state where no pulses are generated. The other two fixed points emerge at a critical feedback gain $\gamma_c=0.73$, exist for $\gamma>\gamma_c$, and are both negative. The fixed point with the smaller magnitude, denoted by $x_u^*$, is unstable, while the fixed point with the greater magnitude, denoted by $x_{s2}^*$, is stable. It corresponds to a periodic pulsating state with amplitude $x_{s2}^*$. Thus, the critical perturbation size is given by $|x_u^*|$ because perturbation amplitudes greater than this value will grow in time towards the stable fixed point $x_{s2}^*$.  For $\gamma>1$, where $|x_u^*| \ll 1$, a very good approximation is given by $x_u^* \approx -1/(\gamma d^2)$. 

We determine $x_u^*$ numerically from map (\ref{eqn:map} and convert to physical units (see Fig.\ \ref{fig:comparison}b).  We also determine $x_u^*$ from a numerical simulation of Eqs.\ (\ref{eqn:model}) and (\ref{eqn:dimensionless2}). It is seen that the agreement between the experiments, and predictions of the mapping and simulations are very good.  Most importantly, it is seen that the minimum perturbation size decreased as a function of $\gamma$, implying that noise will eventually destabilize the fixed point for sufficiently large $\gamma$.  A similar procedure can be used to determine the threshold gain $\gamma_{th}$ required to destabilize the fixed point $x_{s1}^*=0$ for a given noise intensity $D=\sqrt{2<x^2>}$.  At threshold, $<x^2>=|x^*_u|^2$, which yields $\gamma_{th} \simeq 1/[d \sin (Dd/\sqrt{2}-2m)]$.

\begin{figure}[h,t,b]
\includegraphics[scale=1]{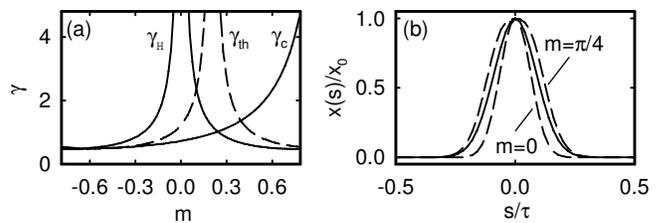}
\caption{\label{fig:stabilityfigure} (a) Instability thresholds using $D=0.28$ in $\gamma_{th}$. (b) Temporal evolution of an initial Gaussian pulse (solid line) after one iteration of the continuous mapping for $m=0$ with $\gamma = 1.23$ and $x_0=-0.2$ (inner dashed line), and for $m=\pi/4$ with $\gamma=0.71$ and $x_0=-0.35$ (outer dashed line).} 
\end{figure}

For the whole range of $m$, $\gamma_c$ continues to indicate the birth of two fixed points, which we determine numerically from map (\ref{eqn:map}) and display in Fig.\ \ref{fig:stabilityfigure}a.  Pulsing behavior is possible for $\gamma > \gamma_c$.  There is a strong asymmetry in $\gamma_c$ about $m=0$, indicating that pulsing behavior is least likely around $m=\pi/4$.  Also shown in Fig.\ \ref{fig:stabilityfigure}a are $\gamma_H$ and $\gamma_{th}$ for one value of the noise intensity.  For $\gamma_{th} < \gamma_H$ ($\sim -\pi/4 < m \lesssim 0.1$), we predict that the fixed point $x_{s1}^*=0$ tends to be destabilized by a pulsing instability.  For $\gamma_H < \gamma_{th}$, $\gamma \sim \gamma_H$, and small noise, we predict that the fixed point tends to be destabilized by the Hopf bifurcation determined from the linear stability analysis.  Thus, we predict that the fixed point will be unstable for $\gamma > \min[\gamma_H,\gamma_{th}]$.  We see that there is qualitative agreement between $\min[\gamma_H,\gamma_{th}]$ shown in Fig.\ \ref{fig:stabilityfigure}a and the high-noise experimental measurements (Fig.\ \ref{fig:gamma_th}b).

Analysis of the map only gives information about the pulse peak amplitude; information about changes in the pulse shape is predicted by the continuous mapping (\ref{eqn:cmap}).  In particular, we take the intial perturbation to be a Gaussian pulse centered at $s=0$, shown in Fig.\ \ref{fig:stabilityfigure}b.  For $m=0$, we find that the pulse remains symmetric and undergoes substantial pulse compression after one round trip through the loop (inner dashed curve).  This pulse compression continues each iterate through the loop, eventually, becoming so short that our asymptotic analysis breaks down when the pulse spectrum exceeds the bandpass filter width.  Thus, we predict that perturbations tend to produce ultrafast pulses whose spectrum fills the available device bandwidth.  Strong frequency mixing, known to occur in time-delay systems, then gives rise to pulse-to-pulse coupling and chaos \cite{Larger2009}. For $m=\pi/4$, the pulse also remains symmetric, but the pulse width expands in a round-trip through the loop.  Thus, short-pulse perturbations tend to smooth out and the system will tend to display more sinusoidal behavior near the instability threshold.  For other values of $m$, the pulses become asymmetric and pulse compression (expansion) occurs when $\gamma_c < \gamma_H$ ($\gamma_H < \gamma_c$.) 
\begin{figure}[t]
\includegraphics[scale=1]{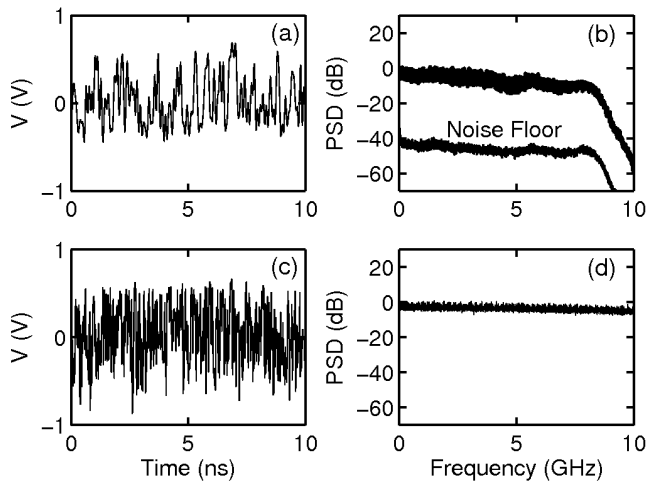}
\caption{\label{fig:bbchaos} The experimental time series (a) and power spectral density (b) of the broadband chaotic behavior in the physical system for $m=0$ and $\gamma = 4.80$ (upper trace). The power spectral density of the noise floor obtained for $m=0$ and $\gamma=4.30$ (lower trace) is also shown. Numerical time series (c) and power spectral density (d) for $m=0$ and $\gamma=4.80$.}
\end{figure}
In the experiment, we observe that, in general, transient behavior similar to that in Fig.\ \ref{fig:comparison}a eventually gives way to chaotic behavior such as the one shown in Fig.\ \ref{fig:bbchaos}a for the case $m=0$ and $\gamma=4.80$, which is just above the noise-induced threshold seen in Fig.\ \ref{fig:gamma_th}.  Figure\ \ref{fig:bbchaos}b shows the one-sided power spectral density (PSD) of the experimental chaotic time series with a resolution bandwidth of 8 MHz.  Interestingly, the power spectrum is essentially `featureless' - flat up to the cutoff frequency of the oscilloscope used to measure the dynamics (8 GHz).  In greater detail, the spectrum is contained with a range of 15 dB with a standard deviation of 3 dB for frequencies below 8 GHz.  Our observation is consistent with our analysis of the continuous map above, and indicates that all frequencies contribute with approximately equal strength and that there are no weakly unstable periodic orbits embedded in the strange attractor.  This behavior contrasts with most other chaotic systems where numerous large peaks appear in the power spectrum.

We compare our results to the case where $\gamma=4.30$, which is just below the threshold for noise-induced instability.  As seen in Fig.\ \ref{fig:bbchaos}b, the power spectral density is at least 40 dB below that observed when the oscillator is in the broadband chaotic state and is consistent with the noise-floor of the overall system.  Furthermore, the noise floor is contained within a range of 18 dB with a standard deviation of 2 dB, indicating that the chaotic spectrum is nearly as featureless as the spectrum of the system noise. Also, once the system exhibits broadband chaos at $m=0$ for sufficiently high gain, the dynamics are not sensitive to changing the bias point by 1 V either way. Figures \ref{fig:bbchaos}c,d show similar broadband behavior of the numerical solution of Eqs.\ (1) and (2). 

%%In summary, we have investigated the dynamics of an opto-electronic oscillator operated in a regime where the quiescent state is expected to be stable.  Through experiments with the physical system and different theoretical approaches, we find that the fixed point is destabilized by a pulsating instability in the vicinity of $m=0$, which causes the system to transition to a coexisting chaotic state.  Such an instability and coexisting chaotic state may have important implications for understanding the stability of other time-delay dynamical systems and may find use, for example, in private chaos communication \cite{LargerNature2005} or chaotic lidar \cite{Lin2004}. 

KEC, LI, ZG, and DJG gratefully acknowledge the financial support of the US Office of Naval Research under award \#N00014-07-1-0734 and the US Army Research Office under grant \#W911NF-05-1-0228.  ES thanks Duke University for their kind hospitality, and acknowledges partial support by the DFG in the framework of Sfb555.

\bibliography{OEchaos}

\end{document}